\documentclass{jpsj2}
\title{Low Temperature Specific Heat of Dy$_2$Ti$_2$O$_7$ in the Kagome Ice State}

\author{Ryuji \textsc{Higashinaka}$^{1}$, Hideto \textsc{Fukazawa}$^{1}$\thanks{Present address: Graduate School of Science and Technology, Chiba University, Chiba 263-8522, Japan}, Kazuhiko \textsc{Deguchi}$^{1}$ and Yoshiteru \textsc{Maeno}$^{1,2}$}

\inst{$^{1}$Department of Physics, Kyoto University, Kyoto 606-8502, Japan \\
$^{2}$International Innovation Center, Kyoto University, Kyoto 606-8501, Japan}

\abst{We report the specific heat of single crystals of the spin ice compound Dy$_2$Ti$_2$O$_7$ at temperatures down to 100 mK in the so-called Kagome ice state. In our previous paper, we showed the anisotropic release of residual entropy in different magnetic field directions and reported new residual entropy associated with spin frustration in the Kagome slab for field in the [111] direction. In this paper, we confirm the first-order phase transition line in the field-temperature phase diagram and the presence of a critical point at (0.98 T, 400 mK), previously reported from the magnetization and specific-heat data. We newly found another peak in the specific heat at 1.25 T below 0.3 K. One possible explanation for the state between 1 T and 1.25 T is the coexistence of states with different spin configurations including the 2-in 2-out one (Kagome ice state), the 1-in 3-out state (ordered state) and paramagnetic one (free-spin state).}

\kword{pyrochlore, Spin Ice, Kagome Ice, geometrical frustration, specific heat}

\begin{document}
\maketitle

\section{Introduction}
Geometrically frustrated systems show various prominent properties, such as spin ice, quantum spin liquid, anomalous Hall effect, etc \cite{Harris97,RamirezN,Canals98,Taguchi01}. Among compounds realizing such systems, pyrochlore oxides $A_{2}B_{2}$O$_7$ have been extensively studied \cite{Review,Snyder03}. Among pyrochlore oxides, Ho$_{2}$Ti$_{2}$O$_7$ \cite{Harris97}, Dy$_{2}$Ti$_{2}$O$_7$ \cite{RamirezN} and Ho$_{2}$Sn$_{2}$O$_7$ \cite{Matsuhira01} are known to exhibit spin ice behavior: Without long-range ordering of the rare-earth magnetic moments, these materials maintain macroscopically degenerate state down to low temperatures \cite{Review}. \\

\begin{figure}[btp]
\includegraphics[width=0.8\linewidth,clip]{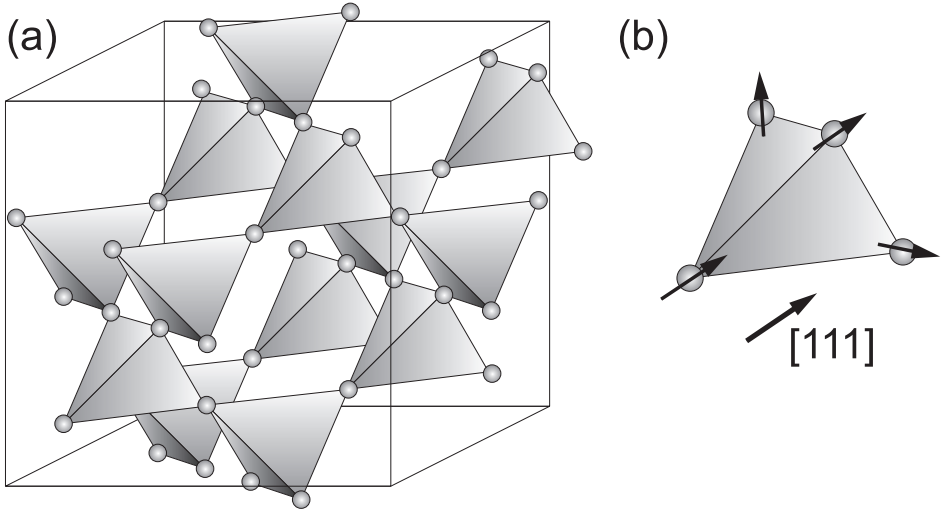}
\caption{\label{Pyrochlore} (a) The pyrochlore lattice: Both $A$ and $B$ sites form the pyrochlore lattice in pyrochlore oxides $A_{2}B_{2}$O$_7$. (b) The stable spin configuration (1-in 3-out) on a single tetrahedron in high field along the [111] direction. Circles depict $A$-site ions with Ising spins. Zeeman interaction in the [111] field direction competes with the spin-spin interaction that stabilizes a 2-in 2-out configuration.}
\end{figure}

\quad In pyrochlore compounds, both $A$-site and $B$-site ions constitute a corner-shared tetrahedral network (the pyrochlore lattice (Fig. 1(a))). Because of the crystal-field effect, some of the magnetic moments of the $A$-site ions such as Dy$^{3+}$ and Ho$^{3+}$ have Ising anisotropy along the local $\langle$111$\rangle$ direction, which points to the center of a tetrahedron from each vertex. Because of the effective nearest neighbor interaction being ferromagnetic owing to the dominant dipolar interaction, the ground state of a single tetrahedron is governed by the `ice rule' in which two spins point inward and the other two point outward (2-in 2-out). Such a configuration on each tetrahedron leads to a macroscopic degeneracy in the pyrochlore lattice and gives rise to Paluing's residual entropy of (1/2)$R$ln(3/2) \cite{Pauling45}. \\
\quad Owing to the Ising anisotropy, the spin responses to magnetic fields are very anisotropic. For polycrystalline samples, it is reported that there are specific-heat peaks at field-independent temperatures of 0.34 K, 0.47 K and 1.12 K; it was speculated that these peaks may be due to ordering of spins with their Ising axes perpendicular to the field \cite{RamirezN}. Moreover, owing to the difference in the stable spin configurations in fields along different directions and associated difference in the frustration dimensionality, the process of releasing the residual zero-point entropy should be qualitatively different. Magnetization measurements using single crystalline Dy$_{2}$Ti$_{2}$O$_7$ revealed such anisotropic spin responses \cite{Fukazawa02,Sakakibara03}. From that study we found that the state in the field along the [111] direction is qualitatively different from those along the other directions [100] and [110]. In particular the state in the [111] field direction exhibits a new value of the residual entropy, because the lattice on which spins are frustrated changes from three-dimensional (3D) pyrochlore lattice to two-dimensional (2D) Kagome lattice with the ice rule constraint. We call this state as the ``Kagome ice state'' \cite{Matsuhira02,Wills02}. \\
\quad In the [111] field direction, Zeeman interaction originating from the external magnetic field favors the 1-in 3-out spin configuration on each tetrahedron (Fig. 1(b)), whereas the ice rule originating from spin-spin interaction tends to stabilize the 2-in 2-out state. One of the spins on each tetrahedron is parallel to the magnetic field and the components parallel to magnetic field of the other three spins are equivalent. The angle $\theta$ between these three spins and the magnetic field gives cos$\theta$ = 1/3; the Zeeman interaction for these spins is one third of that for the field-parallel spin. Therefore, in certain field range the direction of parallel spin is uniquely decided by the Zeeman interaction, whereas the other three still remain frustrated by competition between the Zeeman interaction and the spin-spin interaction. Viewed from the [111] direction, pyrochlore lattice consists of stacking of a triangular lattice and a Kagome lattice. Spins parallel to the magnetic field form the triangular lattice and the other frustrated spins form the Kagome lattice. Because the spins on the Kagome lattice are frustrated with the ice-rule constraint, this state has a different value of residual entropy from that of the spin ice state. From specific heat measurements down to 350 mK, the residual entropy of Kagome ice state was estimated as 0.44 $\pm$ 0.12 J/mol-Dy K by our group \cite{Higashin03} and 0.65 J/mol-Dy K by Hiroi and Matsuhira $et$ $al$. \cite{Matsuhira02,Hiroi03}. From the magnetization of the latter group, it was estimated as 0.5 $\pm$ 0.15 J/mol-Dy K \cite{Sakakibara03}. From a theoretical point of view, the Kagome ice state maps onto a hardcore dimer model on the hexagonal lattice. The residual entropy for this model can be calculated exactly and is 0.6718 J/mol-Dy K \cite{Moessner01,Udagawa02}. \\
\quad Ramirez $et$ $al$. reported the specific heat of Dy$_2$Ti$_2$O$_7$ down to 0.2 K \cite{RamirezN}. However, their samples were polycrystals and they could not extract the properties of the Kagome ice state in the field along the [111] direction. To the best of our knowledge, there is no detailed report of the specific heat measurement below 0.35 K in the Kagome ice state \cite{HigashinHFM2003}. In this paper, we report the specific heat of single-crystalline Dy$_2$Ti$_2$O$_7$ down to 100 mK in order to examine the detailed nature of the Kagome ice state at low temperatures. In addition to the specific-heat peak at 0.98 T also present at higher temperatures, we found a new broad peak around 1.25 T emerging below 0.3 K. We discuss the origin of this additional peak. \\

\section{Experimental}
Single crystals of Dy$_2$Ti$_2$O$_7$ used in this work were grown by a floating zone method \cite{Fukazawa02}. We measured the specific heat between 0.1 and 3 K and in fields up to 2 T by a relaxation method using a self-made calorimeter with a dilution refrigerator (Oxford Instrument model Kelvinox25). In order to attain accurate field alignment along the [111] direction, we used a single-axis sample rotator \cite{NishiZaki00}. The sample size was approximately $2.0 \times 2.0 \times 0.06$ mm$^3$ and the mass was 1.6 mg. The [111] direction lies in the surface plane of the plate-like sample in order to reduce the demagnetization effect. In fact, the demagnetization factor is estimated to be as small as $N = 0.03$; we do not make corrections in the data presented below. We evaluated the specific heat of the addenda from the specific heat measurement of a single crystal of aluminum. \\

\begin{figure}[btp]
\includegraphics[width=0.8\linewidth]{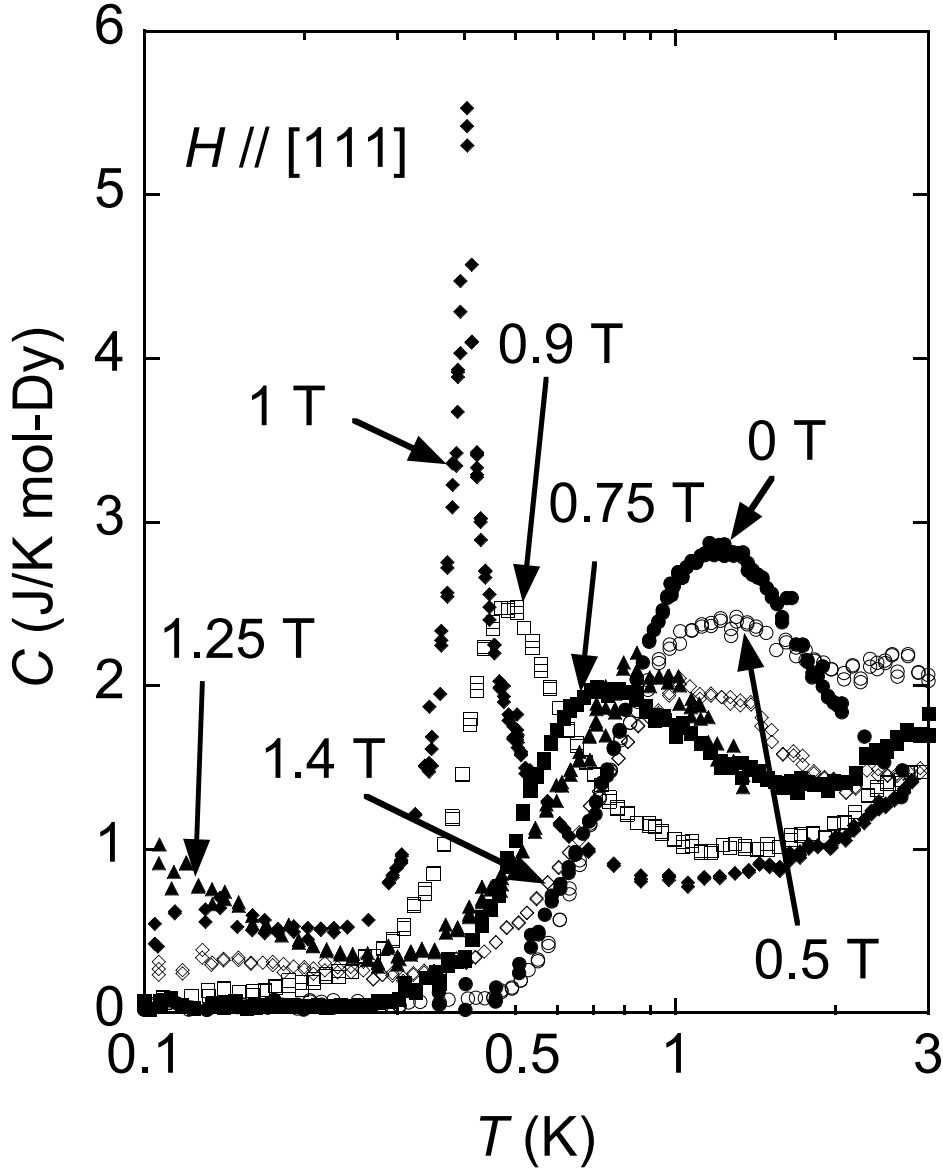}
\caption{\label{TdepC} Temperature dependence of the specific heat of Dy$_2$Ti$_2$O$_7$ at various magnetic fields along the [111] direction down to 100 mK. At 0 T, there is only one peak indicating spin freezing. However, there are two peaks at 0.5 and 0.75 T. The higher one (peak 1) is originating from the spins parallel to the magnetic field and the lower one (peak 2) is originating from the frustrated spins as discussed in text. The higher peak above 0.9 T exists above 3 K.}
\end{figure}

\quad In Fig. 2, we show the temperature dependence of the specific heat in various magnetic fields. In zero field, there is a broad peak at 1.23 K, characteristic of short-range ordering in a tetrahedron by spin freezing. Below this temperature the local 2-in 2-out configuration is stabilized, but does not lead to a long-range ordering. The dipolar spin-ice model predicted that if thermalization process is efficient there should be a sharp peak originating from first-order transition at 0.18 K in the temperature dependence of specific heat at 0 T \cite{Melko01}. However, as reported in our previous paper, we did not detect the theoretically predicted first-order transition at 0.18 K by ac susceptibility measurement down to 60 mK \cite{Fukazawa02}. In the present study, we confirm the absence of a transition in specific heat at 0 T down to 100 mK. \\
\quad In this compound, the spin-spin interaction $J_{{\rm eff}} (\equiv D_{{\rm nn}} + J_{{\rm nn}})$ is estimated to be 1.11 K, where $D_{{\rm nn}}$ (= 2.35 K) and $J_{{\rm nn}}$ (= -1.24 K) represent the dipolar interaction and exchange interaction between nearest neighbors \cite{Hertog00}. The energy difference between in and out spins originating from spin-spin interaction is represented as 4$J_{{\rm eff}}$ per $A$ site spin surrounded by six nearest-neighbor spins. The Zeeman energy is $E_{{\rm Z}}(\theta) = g_{J}J\mu_{{\rm B}}(\mu_{0}H) \times {\rm cos}\theta$, where $\theta$ is the angle between the local Ising axis and the field direction, $g_{J}$ is Lande's $g$ factor, $J$ is a total angular momentum, and $\mu_{{\rm B}}$ is the Bohr magneton. For a Dy$^{3+}$ spin, $g_{J}$ = 4/3 and $J$ = 15/2. The energy difference from this interaction is represented as $2E_{{\rm Z}}$ and for the frustrated spins on Kagome lattice and the field-parallel spins in the field along the [111] direction, these values are represented as 2$E_{{\rm Z}}^{{\rm Kagome}} = 2 E_{{\rm Z}}$(109.47\symbol{"17}) = $6.66 \mu_{{\rm B}}(\mu_{0}H)$ and 2$E_{{\rm Z}}^{{\rm parallel}} = 2 E_{{\rm Z}}$(0\symbol{"17}) = $20 \mu_{{\rm B}}(\mu_{0}H)$, respectively. Under the magnetic field (see for example data at 0.5 T in Fig. 2), the peak in the specific heat splits into two. The peak at higher temperature (peak 1) shifts to higher temperature linearly in the external field. This peak is attributable to the spins parallel to the field. Because the Zeeman interaction for the spins parallel to the field is three times greater than the others, the direction of these spins is decided first with increasing field. \\
\quad In contrast, the peak at lower temperature (peak 2) shifts to lower temperature up to 1 T. At this field, there is a sharp peak at around 400 mK. For fields greater than 1 T, the peak 2 shifts to higher temperature and becomes broad. The origin of the peak 2 is attributable to the frustrated three spins. The Zeeman interaction and the spin-spin interaction for these spins compete with each other in the [111] field direction because the former one stabilizes the 1-in 3-out configuration and the latter one stabilizes the 2-in 2-out configuration. At lower external field, the Zeeman interaction is smaller than the spin-spin interaction ($2E_{{\rm Z}}^{{\rm Kagome}} < 4J_{{\rm eff}}$) and the difference between these two interactions decrease with increasing field. Therefore, the characteristic peak due to this origin shifts to lower temperature with increasing field. At higher field ($2E_{{\rm Z}}^{{\rm Kagome}} > 4J_{{\rm eff}}$), the peak shifts to higher temperature. We note that the specific-heat peak at this turnover field at (0.98 T, 0.40 K) is substantially higher and sharper than those in the previous reports \cite{Higashin03,Hiroi03}. This indicates that the alignment in the present experiment is more accurate than that in those reports. Thus the features presented below are considered as the intrinsic properties rather than due to field misalignment.  \\

\begin{figure}[btp]
\includegraphics[width=0.8\linewidth]{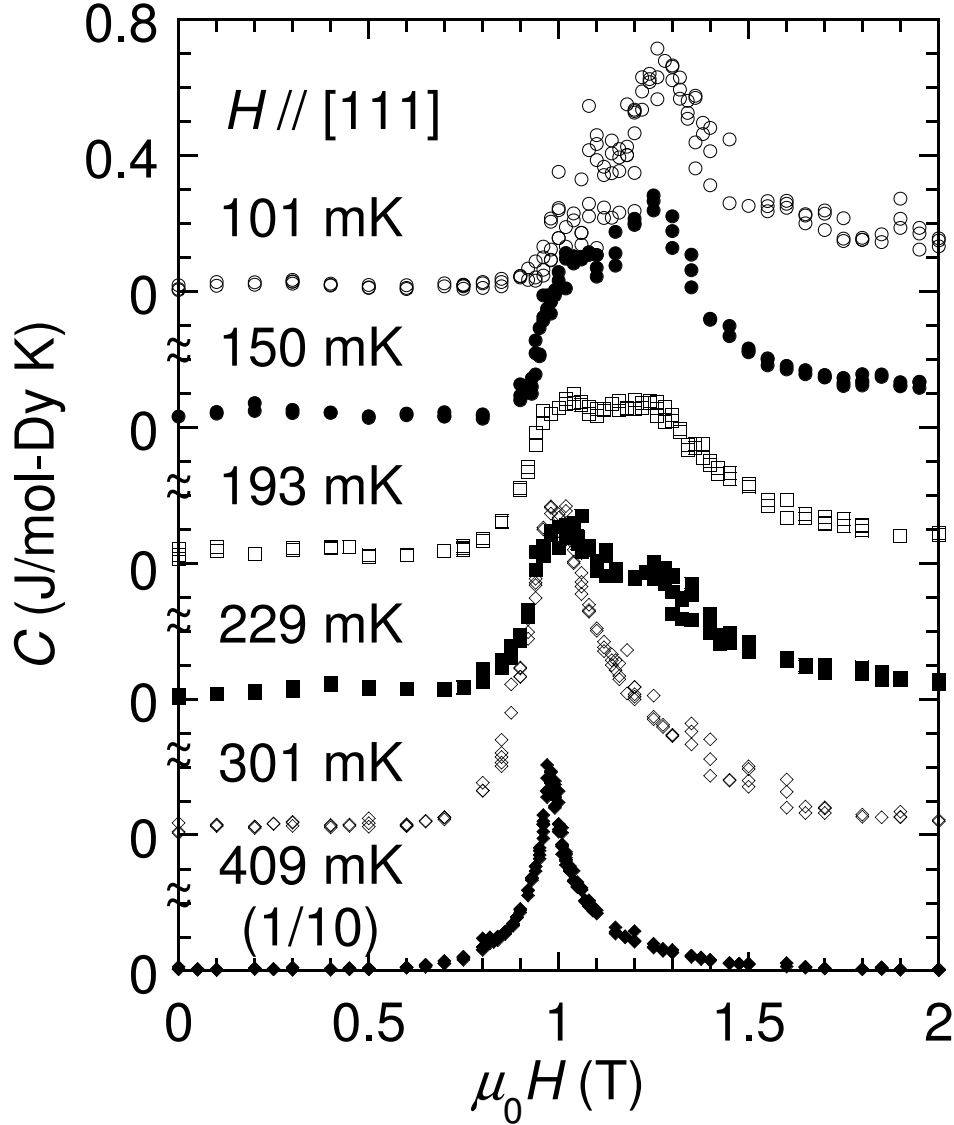}
\caption{\label{HdepC} Field dependence of the specific heat of Dy$_2$Ti$_2$O$_7$ at various temperatures in the [111] field direction. The data at 409 mK is divided by 10.}
\end{figure}

\quad In Fig. 3, we show the field dependence of the specific heat at various temperatures below 0.5 K. Because the specific-heat peak at 409 mK is much higher than those at other temperatures, the values obtained at 409 mK are divided by 10 in Fig. 3. In the temperature range between 230 and 409 mK, there is a temperature-independent peak at 1 T. This field agrees with the field at which the Zeeman interaction is equal to the spin-spin interaction for the frustrated three spins. For these spins, $E_{{\rm Z}}^{{\rm Kagome}}$ is estimated to be 2.24 K at 1 T; thus, the equation $4J_{{\rm eff}} = 2 E_{{\rm Z}}^{{\rm Kagome}}$ is satisfied at $\mu_{0}H$ = 0.99 T, as observed. On the other hand, below 230 mK another peak emerges around 1.25 T. The intensity of the broad 1.25 T peak increases with decreasing temperature. In these experiments, we did not observe any hysteresis. \\

\section{Discussion and Conclusion}

\begin{figure}[btp]
\includegraphics[width=0.8\linewidth]{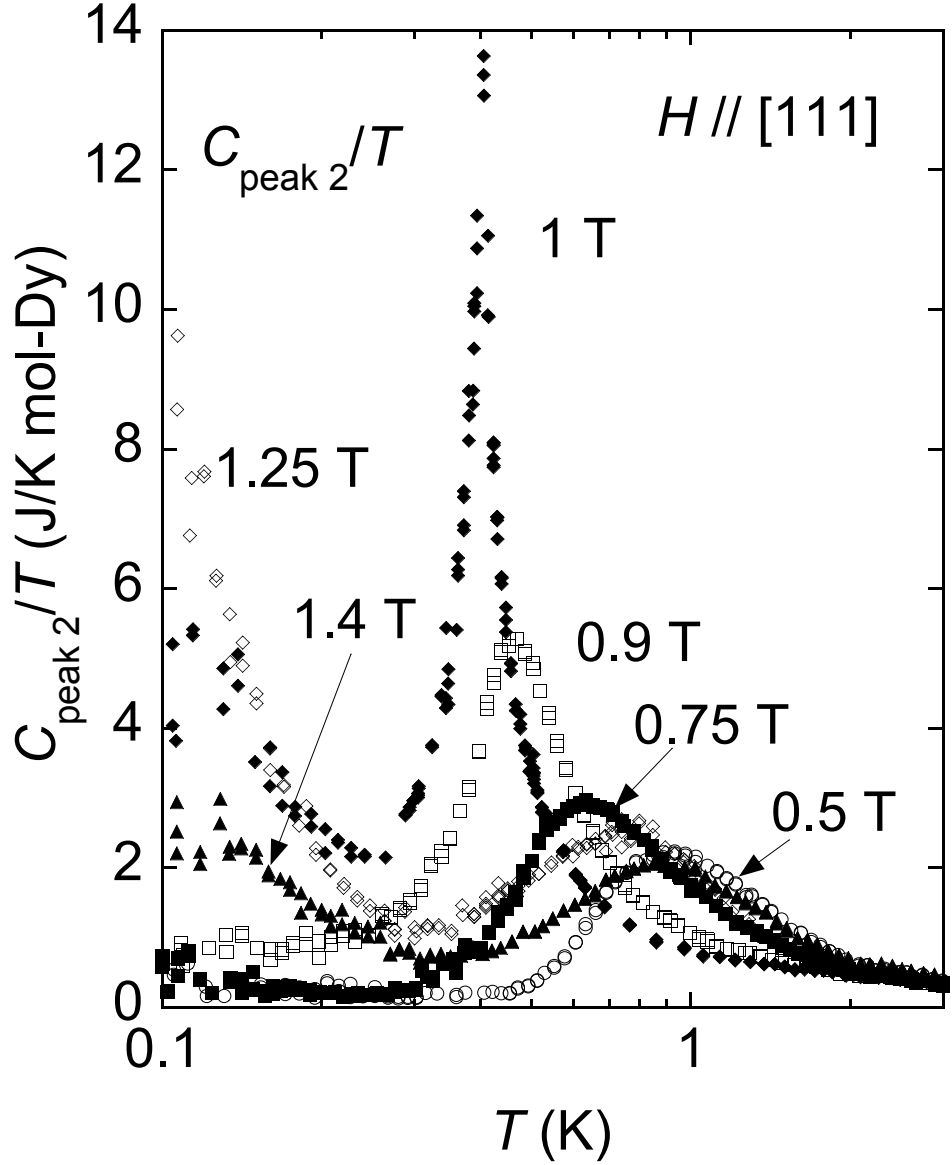}
\caption{\label{TdepC2} Temperature dependence of the specific heat of Dy$_2$Ti$_2$O$_7$ divided by temperature attributable to contribution from frustrated spins at various fields.}
\end{figure}

In Fig. 4, we show the magnetic specific heat divided by temperature attributable to contribution from frustrated spins, defined as $C_{{\rm peak\ 2}}/T = (C_{{\rm total}} - C_{{\rm lattice}})/T - C_{{\rm peak\ 1}}/T$. We estimate the lattice contribution of specific heat using the specific heat of insulating and nonmagnetic Eu$_2$Ti$_2$O$_7$ with comparable mass per formula unit \cite{Higashin03}. In this temperature region, lattice contribution is of the order of $10^{-3}$ of the magnetic one and negligibly small. In the field along the accurate [111] direction (ideal condition), the residual entropy should depend only on the entropy of the frustrated spins. For the contribution from the field-parallel spins we approximate the higher temperature peak as a Schottky peak originating from the sum of the spin-spin interaction and the applied field with some correction,
\begin{equation}
C_{{\rm peak\ 1}} = \frac{1}{4} \frac{N_{{\rm A}}k_{{\rm B}}}{T} \left( \frac{\Delta E}{k_{{\rm B}}T} \right)^{2} \frac{e^{\Delta E/k_{{\rm B}}T}}{(1+e^{\Delta E/k_{{\rm B}}T})^{2}},
\end{equation}
where $N_{{\rm A}}$ is the Avogadro number and $k_{{\rm B}}$ is the Boltzman constant. $\Delta E$ is the energy difference between up and down spins, $\Delta E = 2E_{{\rm Z}}^{{\rm parallel}} \times a + 4J_{{\rm eff}} \times b$. Due mainly to long-range interaction, some corrections are needed for both energy terms. For simplicity, we assume that the two factors are constant and estimate that $a$ = 0.85, $b$ = 0.66 from the fitting of the peak temperatures above 1 K of specific heat measurements \cite{Hiroi03}. \\

\begin{figure}[btp]
\includegraphics[width=0.8\linewidth]{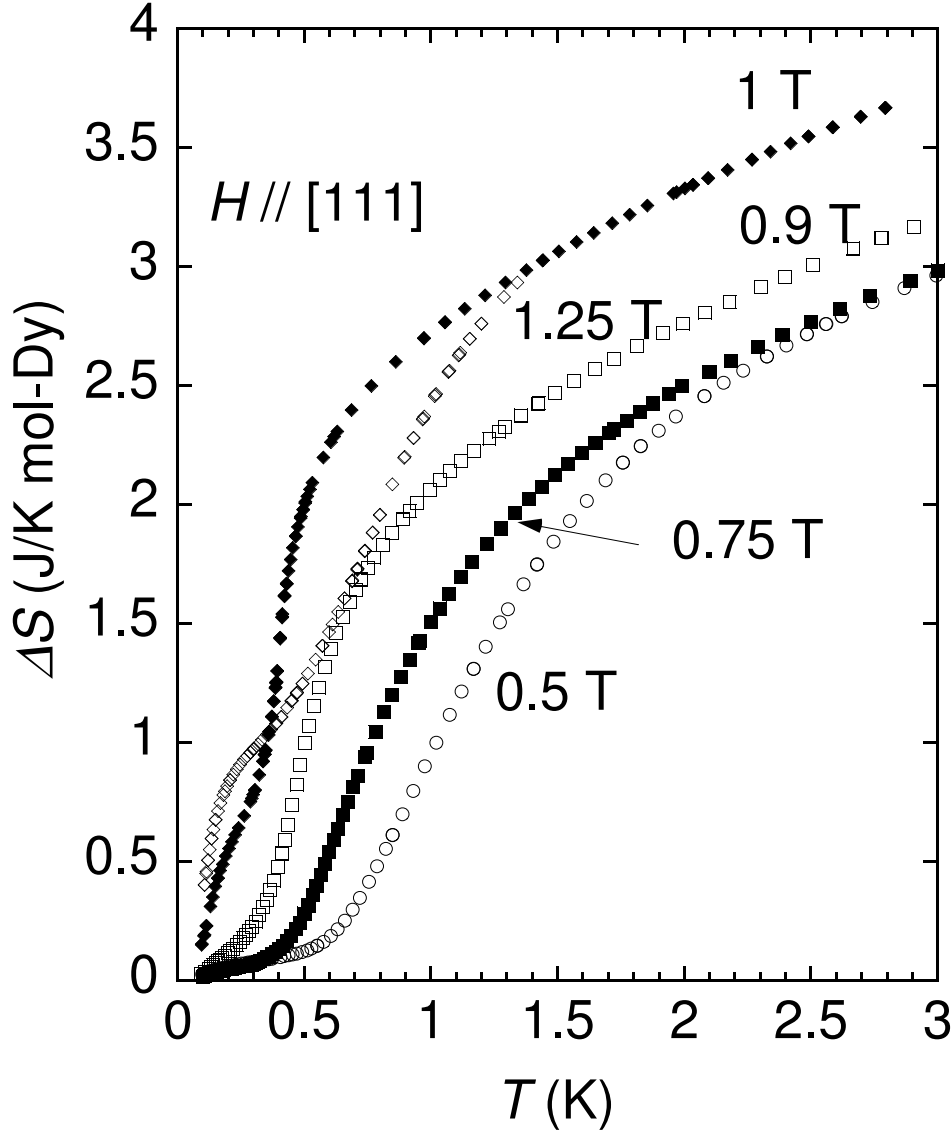}
\caption{\label{TdepS} Temperature dependence of the entropy of Kagome ice state at various fields. The entropy difference at 3 K between the data at 0.5 T and that at 1 T represents the residual entropy of Kagome ice state. Above 1 T, the residual entropy is released below 300 mK.}
\end{figure}

\quad In Fig. 5, we show the temperature dependence of the activation entropy $\Delta S$ attributable to frustrated spins. The entropy at 100 mK, $S$(100 mK), is estimated as $\frac{1}{2} \frac{C({\rm 100\ mK})}{T} \times {\rm 100\ mK}$ by a linear extrapolation of $C/T$ to zero at $T$ = 0. Thus the activation entropy $\Delta S(T)$ for $T$ $\textgreater$ 100 mK is estimated as
\begin{equation}
\Delta S(T) = S({\rm 100\ mK}) + \int^{T}_{{\rm 100\ mK}}\frac{C_{{\rm peak\ 2}}}{T} dT. 
\end{equation}
At 1.4 T, the peak temperature of $C_{{\rm peak\ 2}}$ is higher than that at lower fields; thus, the contribution of peak 2 to the entropy at 3 K is noticeably smaller than that at lower fields. Therefore, we cannot estimate the activation entropy originating from peak 2 to be compared to that at lower fields and we do not show these data in Fig. 5. From this figure, we can estimate the difference between the entropies above 1 T and below 0.9 T. At 2.7 K, the activation entropy $\Delta S$ is 2.82 $\pm$ 0.1 (0.5 T), 2.84 $\pm$ 0.1 (0.75 T), 3.09 $\pm$ 0.1 (0.90 T) and 3.63 $\pm$ 0.1 J/K mol-Dy (1.00 T). The entropy difference between 0.5 T and 1 T is ascribable to the release of the residual entropy of Kagome ice. Around 0.5 T, the ground state retains the residual entropy of Kagome ice. On the other hand, above 1 T the residual entropy is released even below 300 mK. The residual entropy of the Kagome ice state $S_{{\rm KI}}$ is estimated as 0.81 $\pm$ 0.2 J/mol-Dy K in the present study, substantially different from the residual entropy of spin ice state $S_{{\rm SI}}$, 1.68 J/mol-Dy K. The previous results of residual entropy of Kagome ice state are 0.44 $\pm$ 0.12 (our previous one)\cite{Higashin03}, 0.65 \cite{Hiroi03} and 0.5 $\pm$ 0.15 J/K mol-Dy \cite{Sakakibara03}. In our previous experiment, we estimated the entropy below 0.35 K by the procedure similar to that in the present study. Nevertheless, since the alignment of sample was not as good as in the present one, we cannot make a direct comparison of the two sets of the data. On the other hand, in the previous report by Matsuhira $et$ $al$. \cite{Matsuhira02} the definition of the entropy is different from ours. They forced the value of the entropy around 40 K as $R$ln2. However, their estimation of phonon specific heat in the fields was just the fitting $C_{{\rm phonon}} = \alpha T^3$ and in this case it was assumed that additional entropy emerges at high temperature under high magnetic field, possibly from the heat capacity of the addenda. We believe that the differences of the value of the Kagome ice residual entropy between the present and the previous reports are caused mainly by the upturn at low temperature discussed below. In the theoretical reports, only nearest neighbor ferromagnetic interaction is considered \cite{Udagawa02,Moessner01}. In real system, the long-range dipolar interaction is also important, and this should tend to reduce the residual entropy. Thus the theoretical value should give the upper limit of $S_{\rm KI}$. Although not contradictory within experimental uncertainty, however, the present value is also somewhat greater than these theoretical predictions. \\

\begin{figure}[btp]
\includegraphics[width=\linewidth]{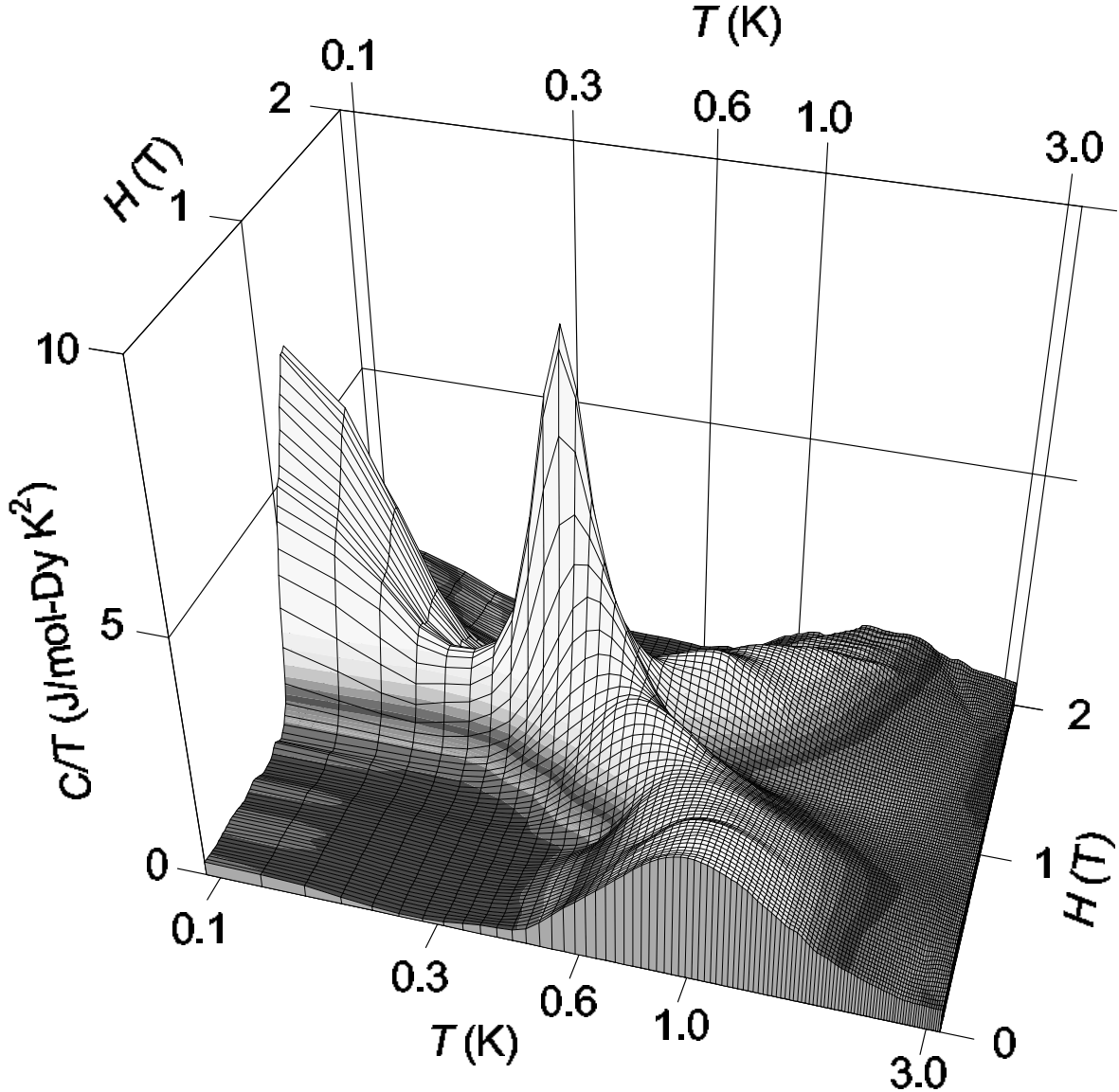}
\caption{\label{3Dplot} The 3D plot of field-temperature dependence of $C/T$ of Dy$_2$Ti$_2$O$_7$ in the field along the [111] direction.}
\end{figure}

\quad We show the 3D plot of the field-temperature dependence of $C/T$ (Fig. 6), as well as the field-temperature phase diagram (Fig. 7) in the field along the [111] direction. In Fig. 7, at zero field below the peak temperature (Fig. 7 A) the spin ice state is realized and the ground state has the residual entropy $S_{{\rm residual}} = S_{{\rm SI}}$. Above this crossover temperature (Fig. 7 B) all the spins are thermally fluctuating and direct randomly; all-random state is realized. At a certain field range and at low temperatures (Fig. 7 C), directions of parallel spins are fixed and Kagome ice state is realized. The ground state has a different residual entropy $S_{{\rm residual}} = S_{{\rm KI}}$. At higher temperatures (Fig. 7 D) the spins on the Kagome lattice are thermally fluctuating while the spins parallel to the field are still pinned; the 1-in 3-random state is stable. The phase diagram above 350 mK reproduces the previous report \cite{Hiroi03}. \\

\begin{figure}[btp]
\includegraphics[width=0.8\linewidth]{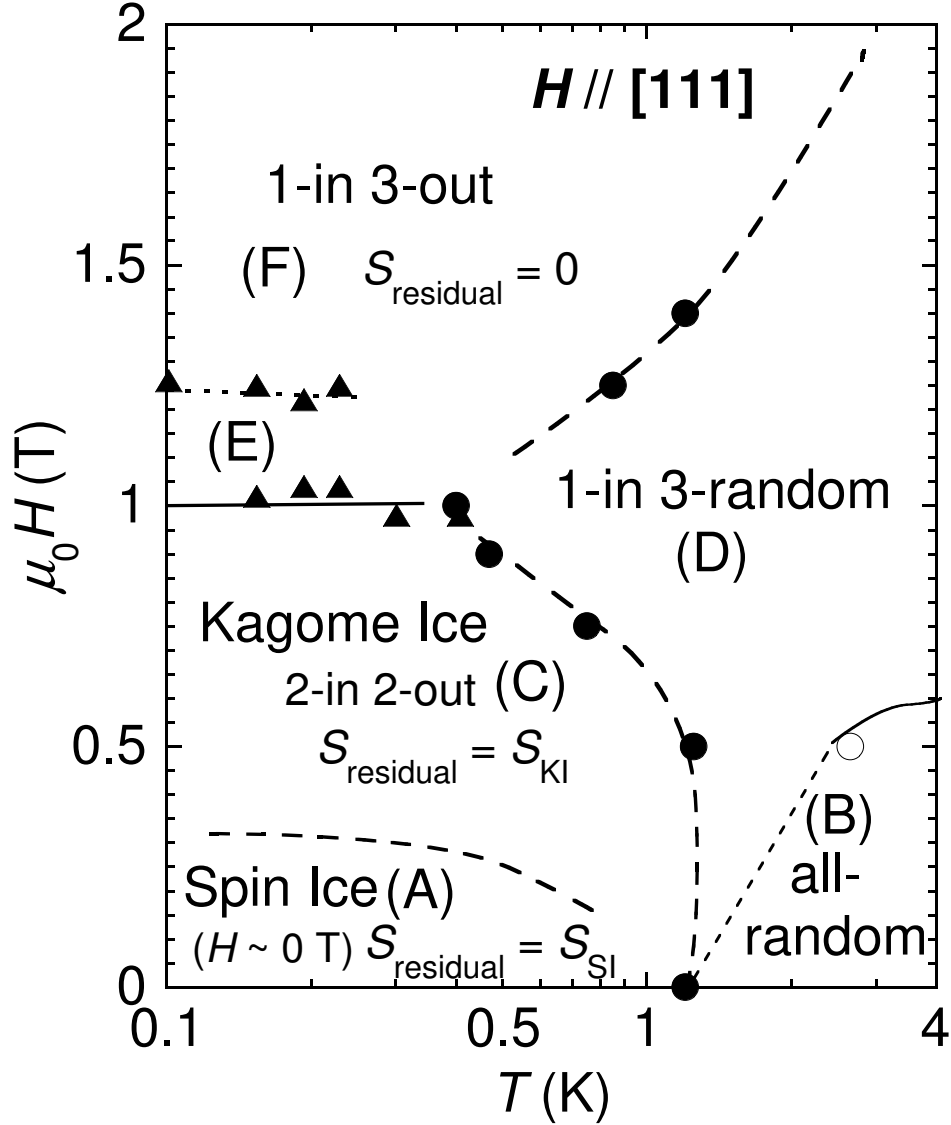}
\caption{\label{PhaseDiagram} The $H$-$T$ phase diagram of Dy$_2$Ti$_2$O$_7$ in the field along the [111] direction. The solid line represents the first-order transition and the dotted lines represent crossover line. The data points correspond to the peak in $C$. }
\end{figure}

\quad At $\mu_{0}H$ = 1 T, there is a transition line nearly parallel to the temperature axis and terminating at a critical end point at $(\mu_{0}H_{{\rm c}}, T_{{\rm c}}) = (0.98\ {\rm T}, 0.40\ {\rm K})$, accompanied by the sharp peak in the specific heat ascribable to the end point of the first order transition. The critical exponent $\alpha$ and $\alpha '$ are represented as
\begin{equation}
C_{H} \propto \left( \frac{T_{{\rm c}}-T}{T_{{\rm c}}} \right) ^{-\alpha '}\ (T\ \textless \ T_{{\rm c}})\ ; \ \left( \frac{T-T_{{\rm c}}}{T_{{\rm c}}} \right) ^{-\alpha}\ (T\ \textgreater \ T_{{\rm c}}).
\end{equation}
For narrow fitting ranges 0.360 K $\textless T \textless$ 0.405 K for $\alpha '$ and 0.413 K $\textless T \textless$ 0.496 K for $\alpha$, we obtain $\alpha '$ = 0.51 and $\alpha$ = 0.38. For a 3D Ising system with common Ising axis, $\alpha '$ is equal to 1/8 to 1/16 and $\alpha$ is equal to about 1/8. Since there are four different Ising axis with frustration structure in this system, we cannot compare these value simply.  \\
\quad From a magnetization experiment, Sakakibara $et$ $al.$ found that the transition is of the first order \cite{Sakakibara03}. Our present experiment was not suitable for observing any latent heat because the phase boundary is nearly parallel to the temperature axis; measurement of magnetocaloric effect is needed to clarify this issue. The phase diagram shown in Fig. 7 is similar to the liquid-gas phase diagram \cite{Sakakibara03}. Because there is no change of the order parameter between the Kagome ice state (Fig. 7 C) and the 1-in 3-out ordered state (Fig. 7 F), the dotted line between Kagome ice state and 1-in 3-random state (Fig. 7 D) and that between 1-in 3-out ordered state and 1-in 3-random state represent crossover lines, not phase boundaries. In fact, we did not observe any sign of phase transitions between the regions C and D or D and F. \\
\quad In addition, a broad peak develops at 1.25 T and below 0.3 K. This suggests another state between 1 and 1.25 T (Fig. 7 E). Above 1.25 T (Fig. 7 F), the 1-in 3-out ordered state is expected to be stable. In previous specific heat measurements, the peak shape of the field dependence of the magnetic specific heat, $C_{{\rm mag}} \equiv C_{{\rm total}} - C_{{\rm lattice}}$ at 0.4 K was clearly asymmetric \cite{Hiroi03}. This asymmetry is consistent with the existence of multiple peaks below this temperature and indeed agrees with our present observation. This means that the residual entropy is not entirely released at the first order transition at 0.98 T, but also released around 1.25 T. One possible explanation of the state between 1 T and 1.25 T is the coexistence of some state with different spin configuration including the 2-in 2-out state (Kagome ice state), the 1-in 3-out one (ordered state) and paramagnetic one (free-spin state). Since the Zeeman interaction nearly compensates the effective nearest-neighbor spin-spin interaction in this region, spins on the Kagome lattice behave just like free-spins except for residual long-range interaction and the system may have a larger residual entropy than that in the Kagome ice state \cite{Zhitomersky,Isakov04}. Since the dipolar interaction is important in this system, the state in this region may not be so simple. This conjecture is consistent with the magnetization of Kagome ice state reported by Sakakibara $et$ $al$ \cite{Sakakibara03}. If Kagome ice state changes to the 1-in 3-out ordered state completely by the first order transition at 0.98 T, the magnetization would have a step-function-like change at this field. The magnetizations indeed exhibit an abrupt change at the low-field side of the transition. However, just above the first order transition, magnetization does not reach the saturated moment expected in the 1-in 3-out ordered state but shows a gradual increase with the field after step-function-like change even at 50 mK \cite{Sakakibara03}. Because this behavior is observed at extremely low temperature 50 mK, the origin must be related to magnetic interaction, not to thermal fluctuation. This indicates that Kagome ice state changes to only an incompletely ordered state at the first-order transition and agrees with our interpretation. If the mixture of different spin configurations exists, the system would retain additional entropy and exhibit the upturn at low temperatures in the specific heat. In relation to this picture, Ogata $et$ $al$. predicted that spin ordering may exist in Kagome ice state when a small transverse field (about 0.04 T) is applied in addition to the field (almost 1 T) along the [111] direction \cite{Ogata}. Since the assumed situation is different from our experiment, their prediction does not seem to be applicable to our result. \\
\quad In conclusion, we measured the specific heat of Dy$_2$Ti$_2$O$_7$ in fields along the [111] direction down to 100 mK. We confirmed the residual entropy of Kagome ice state and found a new peak at 1.25 T below 0.3 K. This peak suggests the mixture of different spin configurations including the 2-in 2-out one (Kagome ice state), the 1-in 3-out state (ordered state) and paramagnetic one (free-spin state) between 1 and 1.25 T and below 400 mK. This interpretation is consistent with previous specific heat and magnetization results \cite{Higashin03,Hiroi03,Sakakibara03}. It should be helpful to examine this conjecture by neutron or NMR experiment in the field along the [111] direction. \\

\section*{Acknowledgment}
\quad We acknowledge helpful discussion with M. Zhitomirsky, S. Nakatsuji, M.J.P. Gingras, H. Tsunetsugu, S. Fujimoto and M. Udagawa. This work was in part supported by the Grant in-Aid for Scientific Research (S) from the Japan Society for Promotion of Science, by the Grant-in-Aid for Scientific Research on Priority Area ``Novel Quantum Phenomena in Transition Metal Oxides'' from the Ministry of Education, Culture, Sports, Science and Technology (MEXT) of Japan, and by a Grant-in-Aid for the 21st Century COE program ``Center for Diversity and Universality in Physics'' from MEXT. H.F. is grateful for the financial support from Japan Society Promotion Science.

\end{document}